\documentclass[preprint,superscriptaddress,citeautoscript]{revtex4}
\usepackage{graphicx}
\setcitestyle{super}
\usepackage[english]{babel}
\usepackage{amsmath}
\usepackage{natbib}		

\begin{document}

\author{Gabriele Bulgarini}
\email{g.bulgarini@tudelft.nl}
\author{Michael E. Reimer}
\altaffiliation{These authors contributed equally to this work}
\author{Mo\"{i}ra Hocevar}
\affiliation{Kavli Institute of Nanoscience, Delft University of Technology, The Netherlands}
\author{Erik P.A.M. Bakkers}
\affiliation{Kavli Institute of Nanoscience, Delft University of Technology, The Netherlands}
\affiliation{Eindhoven University of Technology, The Netherlands}
\author{Leo P. Kouwenhoven}
\author{Valery Zwiller}
\affiliation{Kavli Institute of Nanoscience, Delft University of Technology, The Netherlands}

\keywords{quantum dot; nanowire; single photon detection; photodiode; avalanche; APD}
\title{Avalanche amplification of a single exciton in a semiconductor nanowire}

\begin{abstract}
\textbf{Interfacing single photons and electrons is a crucial ingredient for sharing quantum information between remote solid-state qubits \cite{Zrenner2002,Koppens2006,Press2008,Brunner2009,ZrennerNatPhot,Nadj2010,Benny2011}. Semiconductor nanowires offer the unique possibility to combine optical quantum dots with avalanche photodiodes, thus enabling the conversion of an incoming single photon into a macroscopic current for efficient electrical detection. Currently, millions of excitation events are required to perform electrical read-out of an exciton qubit state \cite{Zrenner2002,ZrennerNatPhot}. Here we demonstrate multiplication of carriers from only a single exciton generated in a quantum dot after tunneling into a nanowire avalanche photodiode. Due to the large amplification of both electrons and holes ($>10^4$), we reduce by four orders of magnitude the number of excitation events required to electrically detect a single exciton generated in a quantum dot. This work represents a significant step towards single-shot electrical read-out and offers a new functionality for on-chip quantum information circuits.
}
\end{abstract}

\maketitle

Semiconductor quantum dots have been proposed as fundamental elements of a quantum information processor \cite{Loss}. Quantum bits (qubits) can be stored in either the spin state of a single electron \cite{Koppens2006,Press2008}, the spin state of a single hole \cite{Brunner2009}, the dark exciton state \cite{Poem2010} or the bright exciton state \cite{Benny2011}. Among these, exciton qubits are particularly attractive for long distance quantum communication through the exciton-to-photon and photon-to-exciton transfer of quantum information \cite{Vrijen2001,Kosaka2009}. Coherent optical control of a single exciton qubit in quantum dots has been demonstrated with both optical \cite{Benny2011} and electrical \cite{Zrenner2002,ZrennerNatPhot} read-out. However, current read-out schemes demand millions of excitation events in order to produce a measurable signal. To reduce the number of excitation events towards single-shot measurements, an internal pre-amplification of the electrical signal would be desirable. Avalanche photodiodes (APD) ensure this high internal gain and generate a macroscopic current in response to a single photon absorbed \cite{Capasso1987}. As a result, single photons can be efficiently detected \cite{Thomas2010}. The avalanche process has been shown at the nano-scale using semiconductor nanowires \cite{LieberNL,Hayden2006,Michael} and carbon nanotubes \cite{Gabor2009}, with sensitivity limited to, at best, approximately 100 photons. Importantly, semiconductor nanowires grown via bottom-up techniques offer an unprecedented material freedom in growing advanced heterostructures \cite{Tomioka2010,Magnus2011} due to reduced strain achieved in the growth of highly mismatched materials \cite{Messing2011}, and can additionally include hetero-structured \cite{MinotNL} or electrically defined \cite{Nadj2010} quantum dots for quantum information processing.

In this work, we have integrated a single quantum dot in the avalanche multiplication region of a nanowire photodiode, which features high internal gain and sensitivity to single photons. By spectrally and spatially separating the absorption region from the multiplication region, we can selectively generate a single exciton in the quantum dot that is efficiently multiplied after tunneling in the nanowire depletion region under an applied electric field. We characterize the nanowire internal gain by photocurrent measurements down to the single photon regime. Finally, we show that only 120 individual excitation events are required to perform electrical read-out of an exciton confined in a single quantum dot.
A scanning electron micrograph of our device containing a single quantum dot embedded in a contacted InP nanowire is depicted in Fig. 1A. The InP nanowire has been doped in-situ during vapor-liquid-solid growth in order to obtain a p-n junction (see Methods). The depletion region of the p-n junction is used to multiply both electrons and holes as they gain enough energy to initiate the avalanche multiplication process. The operating principle of our nanowire avalanche photodiode is shown schematically in Fig. 1B and Fig. 1C. A single photon incident on the device with a frequency of one of the quantum dot energy levels is absorbed and creates a single exciton. Under reverse bias ($V_{sd}\,<\,0$), the electron and hole separate and tunnel into the nanowire depletion region. Next, both the electron and hole accelerate under the applied electric field and once the carriers gain enough energy, additional electron-hole pairs are created by impact ionization. These additional electron-hole pairs can further trigger carrier multiplication and strongly enhance the photocurrent. The final result is that each exciton created in the quantum dot is multiplied into a macroscopic current. A unique feature of InP nanowires is that the impact ionization energy is similar for both electrons and holes: 1.84 and 1.65\,eV, respectively \cite{Pearsall1979}. Both carriers can thus contribute to the avalanche multiplication process and large gains are obtained.
In Fig. 1D we probe the quantum dot and nanowire absorption spectra by tuning the laser excitation wavelength. The measured photocurrent shows a broad absorption peak around 825\,nm, originating from InP band-edge transitions and suggests the presence of both wurtzite and zincblende crystal structures confirmed by transmission electron microscopy \cite{Algra2008}.
Three equally spaced photocurrent peaks are observed at higher excitation wavelengths: at 1007, 986 and 963\,nm. By comparison with typical photoluminescence spectroscopy of single quantum dots in intrinsic nanowires (see Supplementary information), we assign these three peaks to absorption in the quantum dot s-, p-, and d-shell, respectively. From photoluminescence spectroscopy of the device at $V_{sd}\,=\,0$\,V, we confirm that the peak at 1007 nm is due to absorption in the quantum dot ground state (s-shell). The observed shell separation of 26\,meV corresponds to a diameter of $\sim\,$27\,nm according to calculations that assume an in-plane parabolic confinement in the quantum dot and in agreement with the quantum dot size measured with transmission electron microscopy \cite{Reimer2011}. Notably, we observe high photocurrent ($>1$\,nA) for resonant absorption in the quantum dot, which is a direct result of efficient multiplication of charges after tunneling in the nanowire.  \par

We now characterize the photo-response of the nanowire avalanche photodiode. Optical excitation above both InP and InAsP bandgaps is utilized in order to excite the entire nanowire depletion region and measure the resulting multiplication gain.
We use a 120 Hz mechanically chopped continuous-wave laser at wavelength $\lambda$\,=\,532\,nm, that is linearly polarized along the nanowire elongation axis. Inset of Fig. 2A shows the temporal response of the photocurrent (blue line) to the laser trigger (dashed line). We obtain a time constant of 1 ms, which is a direct result of the high resistance of the device (60 M$\Omega$).
Current-voltage characterization of the device is shown in the main panel of Fig. 2A as a function of incident optical power on the sample.
At $V_{sd}$\,=\,0, only the p-n junction built-in electric field separates electron-hole pairs to produce a measurable photocurrent without multiplication gain. Here, we measure a linear dependence for the ratio of charge carriers collected at the contacts to the number of photons absorbed by the nanowire, which has been obtained through an estimation of the absorption efficiency ($\sim\,0.3\,\%$, see Methods). We determine that a single photon absorbed in the nanowire is converted into one electron (hole) worth of current with 96\,$\%$ probability (see Supplementary information). This remarkable efficiency is comparable to state-of-the-art nanowire solar cells \cite{Atwater2010}.
In the reverse bias region ($V_{sd}<0$), the dark current is less than 1\,pA until an avalanche breakdown occurs at $V_{sd}\,=\,-15$\,V, as shown in the Supplementary information. Under illumination, the photocurrent increases rapidly as the applied reverse bias is larger than $V_{sd}\,\sim$\,-1\,V and reaches saturation set by the nanowire total series resistance. The occurrence of a light-induced avalanche at rather low applied bias is due to the use of a small depletion region ($\sim200$\,nm) and the low impact ionization threshold in InP.
This photo-response is reproducible and similar behavior has been observed on six different devices.  \par

In Fig. 2B, we strongly attenuate the excitation pulses for investigating the multiplication gain in the few-photon regime. The device is operated below breakdown voltage at saturation of the avalanche multiplication ($V_{sd}$\,=\,-8\,V) and at a temperature of 40\,K in order to optimize the signal-to-noise ratio of the photocurrent (see Supplementary information). Each experimental data point (black circles) corresponds to the current averaged over 120 excitation pulses and with the dark current subtracted in order to obtain the net photocurrent.
From the linear fit (blue line) to the experimental data, we determine an internal gain, $G_i = 2.3\times10^4$.
This high multiplication factor, combined with the low noise of the device ($\sim$0.2\,pA), enables for the detection of a single photon per excitation pulse. We conclude that, for obtaining such high internal gain in an individual nanowire, both electrons and holes contribute to the avalanche multiplication of charges.
The sensitivity to single photons is confirmed by measuring the photocurrent after individual excitation pulses. In Fig. 2C, the current measured with on average one photon per pulse (circles) is compared to the dark current (crosses). We observe that the absorption of a single photon yields a signal well distinguished from the dark current with a signal-to-noise ratio greater than 2. The detection of single photons from an individual InP nanowire does therefore not require the averaging over multiple excitation events or the use of lock-in amplification. Estimation of the number of photons absorbed per excitation pulse was confirmed by photon counting measurements in the few-photon regime, which are well-described by Poisson statistics. Measurements are described in the Supplementary information.

In the following, we use the single quantum dot as the only absorption region of the nanowire photodiode.
By tuning the excitation energy, we selectively excite only the quantum dot transitions and utilize the nanowire as an efficient multiplication channel for the photo-generated carriers. In Fig. 3A we tune the excitation wavelength to the quantum dot strongest resonance, the p-shell transition at $\lambda=986$\,nm, and measure the photocurrent as a function of the laser spot position with respect to the sample. The photocurrent image is superimposed with a laser reflection image to exclude contribution from Schottky contacts. The photocurrent maximum is reached when the laser spot impinges between the metallic contacts. The same photocurrent maximum position is obtained when exciting the entire nanowire with $\lambda\,=\,532\,$nm, confirming that, as expected, the quantum dot is positioned within the nanowire depletion region.

A comparison of the photocurrent measured as a function of incident light polarization angle for above bandgap excitation ($\lambda$\,=\,532\,nm) and resonant quantum dot p-shell excitation ($\lambda\,=\,986\,$nm) is shown in Fig. 3B. As typically observed in nanowires, the nanowire acts as a dielectric antenna for incoming light and thus photons polarized along the nanowire are favorably absorbed \cite{Wang2001,Kouwen2010}. We observe this antenna effect both for excitation in the entire nanowire active area (crosses) and for selective excitation in the quantum dot (circles). \par
Finally, we estimate the gain obtained after a single photon is resonantly absorbed in the quantum dot p-shell and tunnels into the nanowire multiplication region. Remarkably, we measure a large multiplication gain of $1.3\,\times\,10^4$ (Fig. 3C). As a result, the generation of a single exciton in the quantum dot is detected using a low excitation rate of 120 Hz, where only a single photon is absorbed in the quantum dot per excitation pulse. The observed detection efficiency is four orders of magnitude larger than previously reported on self-assembled quantum dots embedded in diode structures \cite{Zrenner2002} and seven orders of magnitude larger with respect to photodetectors based on a single contacted quantum dot in a nanowire \cite{Kouwen2010}.\par

We have demonstrated that InP nanowires provide an efficient one-dimensional channel for electrical transport, where electrons and holes undergo impact ionization with high probability, thereby achieving large multiplication factors.
The significant gain ($>10^4$) we report for resonant absorption in a single quantum is promising towards single-shot electrical read-out of an exciton qubit state for the transfer of quantum information between flying and stationary qubits \cite{Vrijen2001,Kosaka2009}. The extremely small active area represents the fundamental limit to the device external efficiency, which can be enhanced by light harvesting methods such as plasmonic \cite{Curto2010} and dielectric \cite{Lee2011} antennas. On the other hand, in the present configuration the quantum dot absorption region can be spectrally and spatially separated from the multiplication region. The material freedom available during nanowire growth enables for engineering the quantum dot absorption properties, while conserving single photon sensitivity with a unique sub-wavelength spatial resolution.

\clearpage
\begin{figure}
\includegraphics{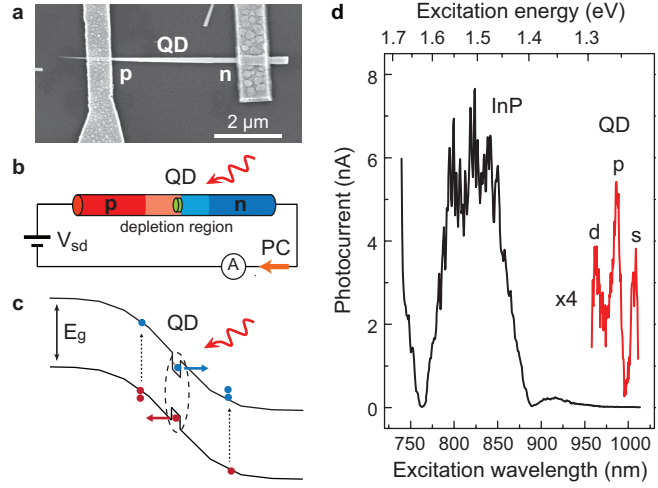}
\caption{A single quantum dot in a nanowire avalanche photodiode. \textbf{a}, Scanning electron micrograph of the nanowire device containing an individual quantum dot (QD). \textbf{b}, The quantum dot is situated within the depletion region of the photodiode, where multiplication of the net photocurrent (PC) is achieved under reverse voltage bias ($V_{sd}$). \textbf{c}, Schematics of carrier multiplication starting from an exciton generated in the quantum dot, followed by tunneling in the nanowire depletion region and an avalanche multiplication of charges through impact ionization driven by the applied electric field. \textbf{d}, Photocurrent spectroscopy obtained by tuning the excitation wavelength with 1 $\mu$W and 20 $\mu$W excitation power for the black and red curve, respectively. In both cases, the device is operated at $V_{sd}\,=\,-2\,$V and the laser is focused on the nanowire depletion region. Band-edge absorption in the nanowire is observed around 825\,nm (black). At longer wavelength, absorption in the quantum dot s-, p- and d-shell is observed at 1007, 986 and 963\,nm, respectively (red).}
\end{figure}

\clearpage
\begin{figure}
\includegraphics{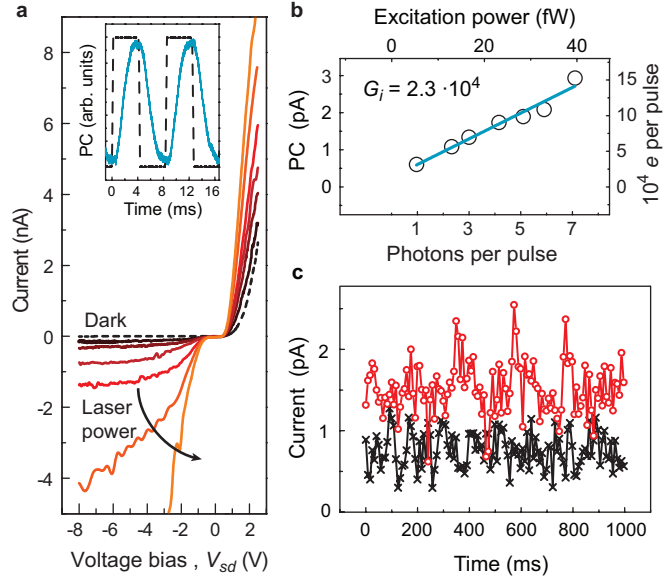}
\caption{Single photon detection with a nanowire photodiode. \textbf{a}, The main panel shows electrical characteristics of the device in absence of illumination (dashed line) and as a function of incident optical power on the sample (100 pW, black - 100 nW, orange) at $\lambda = 532\,$nm (colored curves). The laser is mechanically chopped at 120\,Hz. Inset shows the photocurrent temporal response (blue line) to laser trigger pulses (black dashed line). \textbf{b}, Nanowire photodiode response in avalanche operation at $V_{sd}$\,=\,-8\,V. Dark current is subtracted to obtain the photocurrent (PC). High multiplication gain of $2.3\times10^4$ and the low noise (0.2\,pA) enable the detection of one absorbed photon per excitation pulse, that is further confirmed by measuring after individual laser pulses in \textbf{c}. The absorbtion of on average one photon per pulse yields a signal (circles) well distinguished from the dark current (crosses).}
\end{figure}

\clearpage
\begin{figure}
\includegraphics{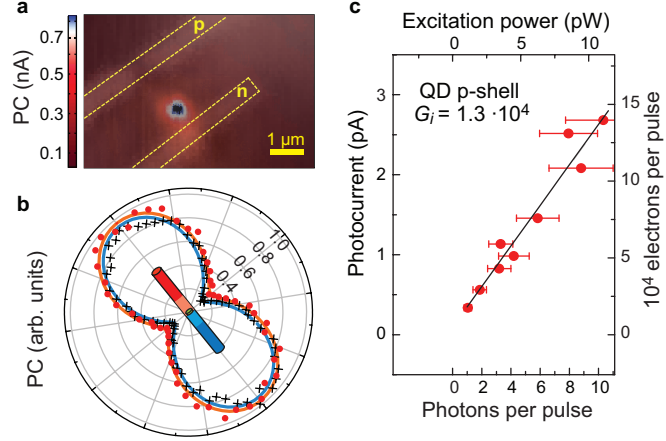}
\caption {Resonant single photon detection in the quantum dot. \textbf{a}, By tuning the excitation wavelength, the quantum dot can be selectively excited and its position is imaged by scanning the laser on the sample with excitation energy resonant to the quantum dot p-shell. The photocurrent (PC) image is superimposed with a laser reflection image to identify the position of the photocurrent maximum with respect to the metallic contacts. As expected, the maximum occurs within the p-n junction where the quantum dot is located. \textbf{b}, Measurement of the linear polarization sensitivity of the nanowire photodiode for excitation at $\lambda\,=\,532\,$nm (blue fit to crosses) and resonant excitation in the quantum dot (red fit to circles). The nanowire acts as a dielectric antenna for resonant optical excitation in the quantum dot, yielding a high polarization sensitivity. \textbf{c}, From the photocurrent response measured at $V_{sd}$\,=\,-8\,V for p-shell excitation, we obtain an average of $1.3\times10^4$ electrons generated per each photon absorbed in the quantum dot. Error bars represent the confidence range in determining the quantum dot height.}
\end{figure}
\clearpage
\noindent\textbf{Methods:}\par

\noindent\textbf{Device fabrication} \par
\noindent InP nanowire p-n junctions containing InAsP quantum dots were grown by vapor-liquid-solid mechanism using gold catalysts of diameter 20 nm. Growth is performed in a metal-organic vapor phase epitaxy (MOVPE) reactor at a pressure of 50 mbar. Tri-methyl-indium, phosphine and arsenic were used as precursors for nanowire and quantum dot growth, while doping was achieved by supplying hydrogen sulfide and diethyl-zinc for n- and p-doping, respectively. The growth is initiated with the n-InP section, followed by the InAsP quantum dot and the p-InP section. The quantum dot features a height of 6 $\pm$ 2 nm and a diameter of approximately 20-30 nm. Subsequently to the growth, nanowires were transferred on a Si substrate covered with 285 nm of SiO$_2$ and contacts were designed by electron beam lithography. The n-side of nanowires was contacted with evaporated Ti/Au (100/10 nm), whereas Ti/Zn/Au (1.5/30/70 nm) was used for p-contacts. Since the p-contact has a Schottky behavior, a rapid thermal annealing at 350 $^\circ$C was consequently performed to induce zinc diffusion from the contact into the nanowire and thus to reduce the Schottky barrier width. After annealing, the typical device series resistance is decreased by three orders of magnitude, to few tens of M$\Omega$. Electrical measurements combined with scanning electron microscopy were performed on homogenously n-doped and p-doped InP nanowires in order to measure doping concentrations: 10$^{18}$ and  10$^{17}$ cm$^{-3}$, respectively. From the measured doping levels, we estimate a p-n junction depletion region width of 140 nm at 0 V, which increases up to 370 nm at -8 V applied bias. \par

\noindent \textbf{Photodiode sensitivity}\par
\noindent In order to estimate the photodetection efficiency of the device, the optical size of the laser beam was measured with photocurrent imaging to first estimate the absorption cross section and subsequently, the absorbed photon flux was obtained utilizing bulk absorption coefficients\cite{AspnesPRL}. A green laser diode ($\lambda$\,=\,532\,nm) was focused to an optical width of 1.2\,$\mu m$ with a 0.75 numerical aperture objective. A Ti:sapphire laser was used for photocurrent spectroscopy and to resonantly excite the quantum dot. At $\lambda$\,=\,986\,nm the optical width of the laser spot was 1.8\,$\mu$m. The photodiode active area was calculated by electrical measurements combined with SEM, assuming a rectangular nanowire cross section and flat surfaces. The resulting absorption efficiency at $\lambda$\,=\,532\,nm is $\sim$\,0.3\,$\%$ at V$_{sd}$\,=\,0\,V and $\sim$\,0.8\,$\%$ at V$_{sd}$\,=\,-8\,V. At this excitation wavelength, both the InP nanowire and the InAsP quantum dot are excited, but due to the large difference in size we can neglect the quantum dot contribution and attribute absorption at $\lambda$\,=\,532\,nm to only the nanowire section. The absorption efficiency obtained under resonant excitation in the quantum dot at $\lambda$\,=\,986\,nm is $\sim$\,0.003\,$\%$. The optical power of the laser was measured with a calibrated power meter and attenuated with neutral density filters. Laser transmission through filters and optical elements constituting the optical excitation path was measured to calibrate the optical power impinging on the sample. Photon counting measurements performed in the few-photon regime confirm estimations of the absorbed photon rates. Details are described in the Supplementary information.

\begin{acknowledgments}
The authors would like to thank J. Rarity and S.M. Frolov for useful scientific discussions.
This work was supported by the Netherlands Organization for Scientific Research (NWO), Dutch Organization for Fundamental Research on Matter (FOM), European Research Council and DARPA QUEST grant.
\end{acknowledgments}

\clearpage 
\textbf{\large{Supplementary information}}
\vspace{0.5cm}

\noindent \section*{Device fabrication and characterization}

\noindent Electrical measurements combined with scanning electron microscopy were performed on homogenously n-doped and p-doped InP nanowires in order to measure doping concentrations. Contacts to n-doped InP nanowires are ohmic and a doping level of 10$^{18}$ cm$^{-3}$ is obtained from two-terminal resistivity measurements. Due to the Schottky nature of the p-contacts, IV characteristics of homogenously p-doped nanowires are highly non-linear. The resistivity of the nanowire without contact contributions was measured using a four-terminal resistivity measurement. Supplementary Fig. 1A shows a scanning electron micrograph of a single homogenously p-doped InP nanowire with four contacts. The two-terminal electrical characteristic of the p-InP nanowire is shown in Supplementary Fig. 1B, whereas Supplementary Fig. 1C shows the resistivity of the only nanowire without contact contribution measured by driving a current through contacts 1-4 and measuring the voltage drop at contacts 2-3. The measured resistivity $\rho$ = 63 $\Omega\cdot$cm suggests a doping concentration of 10$^{17}$ cm$^{-3}$ for the p-doped nanowires. Bulk InP values were used for both electron and hole mobilities \cite{Siegel}.
From the measured doping levels we estimate a p-n junction depletion region width of 140 nm at 0 V applied bias, which increases up to 370 nm when the nanowire is biased at -8 V. \par

In Supplementary Fig. 2 we analyze the breakdown behavior of the device in dark at room temperature. The black curve exhibits the occurrence of a `hard-knee' abrupt increase of the current at $V_{sd}$ = -15 V, typical of a avalanche breakdown. Subsequently to the breakdown, we slowly sweep the applied bias to $V_{sd}$ = 0 V and measure again the dark characteristic of the device, utilizing the same bias sweep rate and direction. We observe the characteristic of the device to be dramatically changed by the reach of the breakdown voltage. At first, the avalanche breakdown shifts to higher reverse bias and becomes less sharp (red curve) until turning into a smooth Zener-like breakdown (blue curve). We attribute this degradation of the photodiode characteristics to diffusion of dopants and consequent change of the doping profile along the nanowire due to the high applied bias and heat caused by the breakdown current. Thus, in order to preserve the quality of the device, we operate the nanowire photodiode in sub-Geiger mode operation, below the avalanche breakdown voltage.

\noindent \section*{Photon counting measurements}

\noindent Estimations of the absorbed photon rates are confirmed by photon counting measurements in the few-photon regime. We measure the photocurrent after individual excitation pulses and we plot in Supplementary Fig. 3A the statistical distribution of the photocurrent, using a bin size of 0.1 pA, for dark (black) and under laser excitation in the few-photon regime with $n$\,=\,1.00 (red), 2.14 (blue) and 4.29 (green). Here, $n$ represents the estimated average photon rate.
The distribution of the photocurrent under no illumination is fit very well with three equally spaced Gaussian peaks (yellow curves) -- the sum is displayed by the black line. The peak at 0.48 pA is attributed to minority carrier drift across the p-n junction, that constitutes the background current for reverse bias operation of the photodiode. Additionally, we observe a maximum of the photocurrent probability at 0.9 pA and a less pronounced peak at 1.3 pA. These two peaks are attributed to the generation of one and two dark avalanches, respectively. The separation between the background and dark avalanche peak is in very good agreement with the measured internal gain, where $2.3\times10^4\,e$/s corresponds to 0.44 pA. Here, $e$ is the electron charge.
Using a weak laser pulse with on average one photon absorbed per pulse, $n$\,=\,1.00 (red), the maximum of the photocurrent probability increases by a discrete amount of current, 0.4 pA, which is the result of to the contribution of one absorbed photon generating an avalanche multiplication of carriers. As $n$ is further increased, the photocurrent distributions shift to higher values and broadens as a direct result of Poisson statistics. Experimental data are well fitted with a sum of multiple Gaussians (yellow curves, sum in colored lines) at discrete photocurrent steps of 0.4 pA, attributed to the absorption of 1, 2, 3 and more photons.
The number of photons detected within a single laser pulse can be discriminated by the photocurrent amplitude in the few-photon regime down to single photons. The absorption of a single photon yields a discrete photocurrent of 0.4 pA, which agrees both with the dark avalanche amplitude and the measured multiplication gain of the device.
Moreover, by normalizing the area of Gaussian fits in Supplementary Fig. 3A, with the dark count rate probability, we obtain the average detected photon-number at different excitation powers. Measured photon-number probabilities (circles) are fitted in Supplementary Fig. 3B with Poisson distributions (stars) with errors four times below the standard deviation (i.e., $\sqrt{\bar{n}}$). The obtained average detected photon-number $\bar{n}$ deviate from the estimated values $n$ on average by less than 10 $\%$, thus confirming the estimation of the absorbed photon rate. \par

The reported photodetection measurements have been performed at a temperature of 40\,K, where the photocurrent signal-to-noise ratio (SNR) is observed to be maximum, as shown in Supplementary Fig. 4A. The SNR has been measured using green excitation with an optical power of 5.6\,fW incident on the sample, equivalent of an estimated flux of one photon absorbed per excitation pulse at 120 Hz. Cooling down the device below 40\,K results in a reduction of the SNR. This can be explained by dopant freezing which decreases the electric field across the p-n junction. To support this explanation, current-voltage measurements in dark as a function of the temperature are shown in Supplementary Fig. 4B. The forward current of the photodiode presents an abrupt decrease at about 40\,K, which is most likely a result of dopant freezing, thus diminishing the p-n junction electric field and increasing the photodiode resistance. By fitting the dark current-voltage curve at 40\,K with a diode characteristic, we obtain a total series resistance of 60\,M$\Omega$.

\noindent\section*{Photodiode internal quantum efficiency}

\noindent The multiplication gain of the photodiode has been calculated by the ratio of charge carriers collected at the contacts to the number of photons absorbed. This ratio has been normalized by the internal quantum efficiency of the device for obtaining the final multiplication factor. The internal quantum efficiency (IQE) of a photodiode is defined as the ratio between charges to absorbed photons when the internal gain is not active. We measured the internal quantum efficiency of our nanowire photodiode at $V_{sd}$\,=\,0\,V where only the p-n junction built-in electric field separates electron-hole pairs and charge carriers do not gain enough energy to initiate avalanche multiplication. Under illumination, the current-voltage characteristic of the nanowire photodiode shows a negative saturation current ($I_{sc}$). The background drift current of the device is subtracted to the saturation current in order to obtain the net number of charge carriers generated by photon absorption (Supplementary Fig. 5A). We observe a linear dependence of the saturation current with the excitation power, resulting in an internal quantum efficiency of 96\,$\%$ (Supplementary Fig. 5B). Additionally, we observe under illumination an open-circuit voltage (V$_{oc}$ = 0.5 V), which is constant for the investigated range of excitation power.

\noindent \section*{Single quantum dot spectroscopy}

\noindent Photocurrent spectroscopy has been utilized for determining quantum dot absorption transitions. The observed absorption spectra has been compared to typical photoluminescence spectroscopy of individual InAsP quantum dots in intrinsic InP nanowires in order to attribute absorption peaks to transitions in the quantum dot energy levels. Typical photoluminescence spectra as a function of excitation laser power are shown in Supplementary Fig. 6 and exhibit the sequential s-, p-, and d-shell filling as the excitation power is increased. The constant energy separation between observed shells, suggests a 2D parabolic in-plane potential as a valid approximation for quantum dots in nanowires. This approximation has been used to calculate the quantum dot diameter (27 nm).

\clearpage
\begin{figure}[htb]
\centering
	\includegraphics{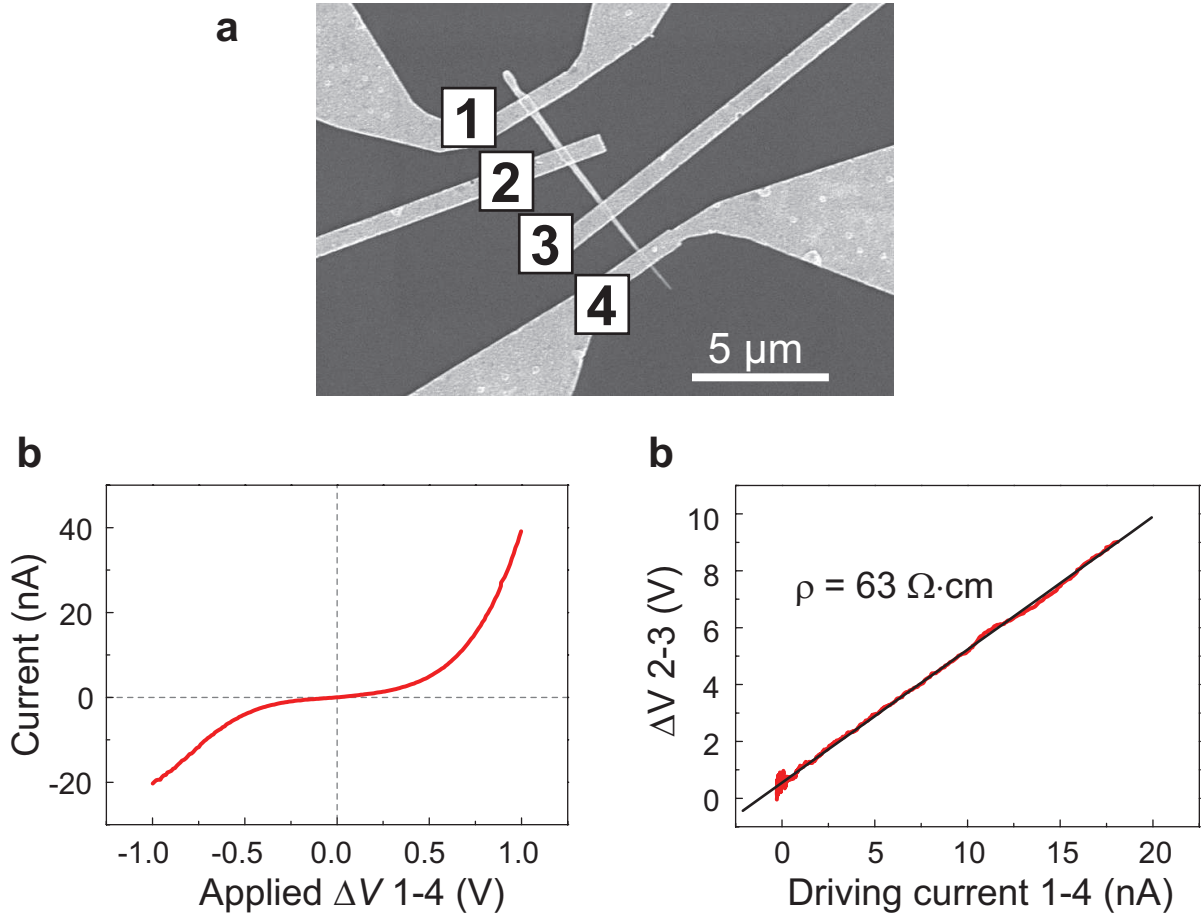}
\caption{Measurement of doping concentration. \textbf{a}, Scanning electron micrograph of a single homogenously p-doped InP nanowire contacted with four contacts. \textbf{b}, The non-linear two terminal I-V characteristic suggests the presence of Schottky contacts. \textbf{c}, The resistivity of the nanowire without the contact contribution is measured with a four-terminal electrical measurement. From the measured resistivity $\rho\,=\,63\,\Omega\cdot$cm, we obtain a doping concentration of 10$^{17}$\,cm$^{-3}$.}
\end{figure}

\clearpage

\begin{figure}[htb]
\centering
	\includegraphics{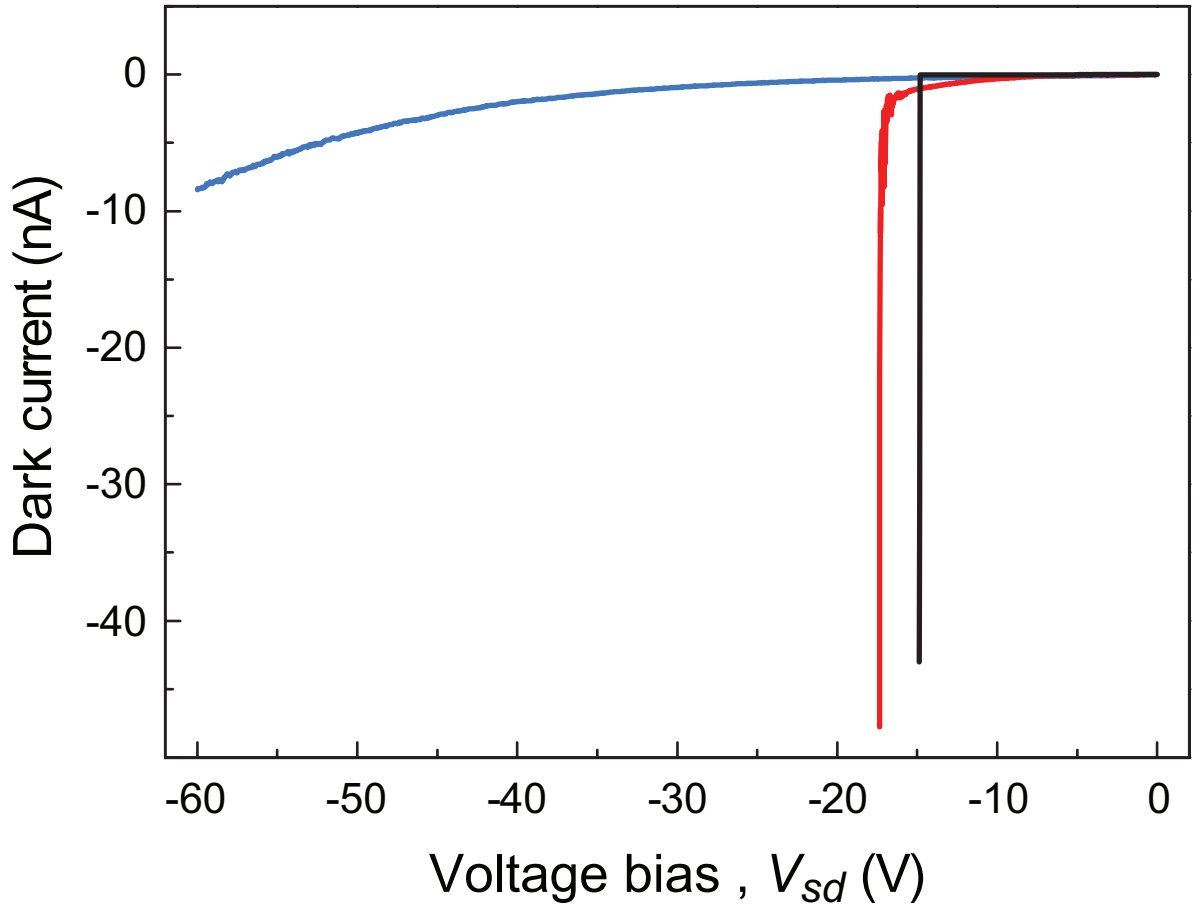}
\caption{Avalanche breakdown in an InP nanowire p-n junction. The black curve shows the photodiode dark current at room temperature, which exhibits the occurrence of a 'hard-knee' avalanche breakdown at $V_{sd}$ = -15 V. The nanowire properties are observed to change after reaching the breakdown. First, the avalanche breakdown shifts to higher reverse bias and becomes less sharp (red curve) and subsequently turns into a smooth Zener-like breakdown (blue curve). The three curves are measured subsequently, under the same bias sweep rate and direction.}
\end{figure}

\clearpage
\begin{figure}[htb]
\centering
	\includegraphics{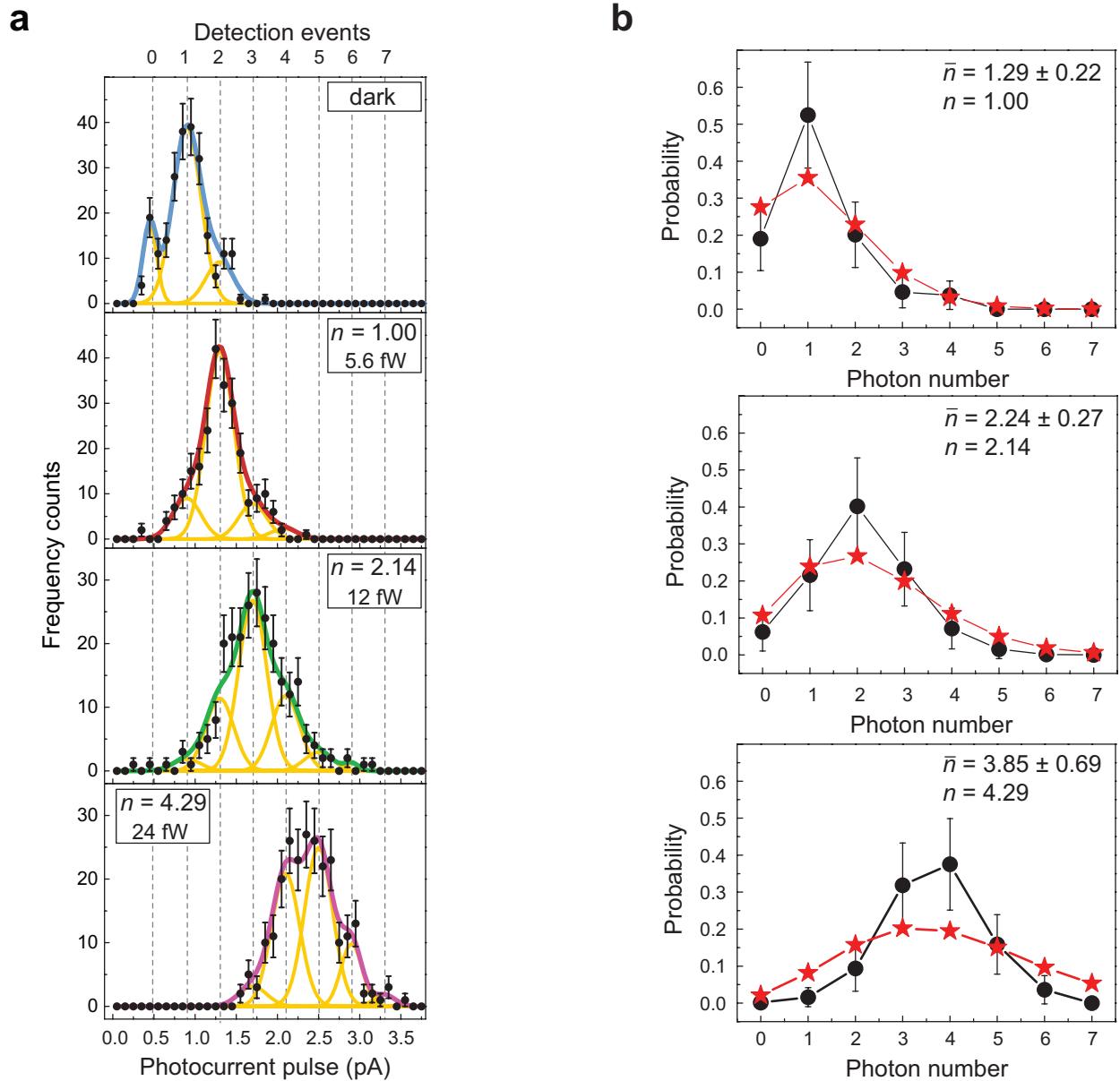}
\caption{Photon counting in the few-photon regime. \textbf{a}, The statistical distribution of photocurrent obtained by measuring after individual excitation pulses with few photon absorbed per pulse, $n$. Experimental data are fitted with Gaussians (yellow curves) equally separated by 0.4 pA, which correspond to the current generated by a single photon absorbed. The sum of Gaussian peaks is showed by the colored curves as a fit to experimental data. \textbf{b}, Poisson distribution (stars) well describes the observed detection probabilities (circles) obtained from the area of fitting Gaussians in \textbf{a}.}

\end{figure}

\clearpage
\begin{figure}[htb]
\centering
	\includegraphics{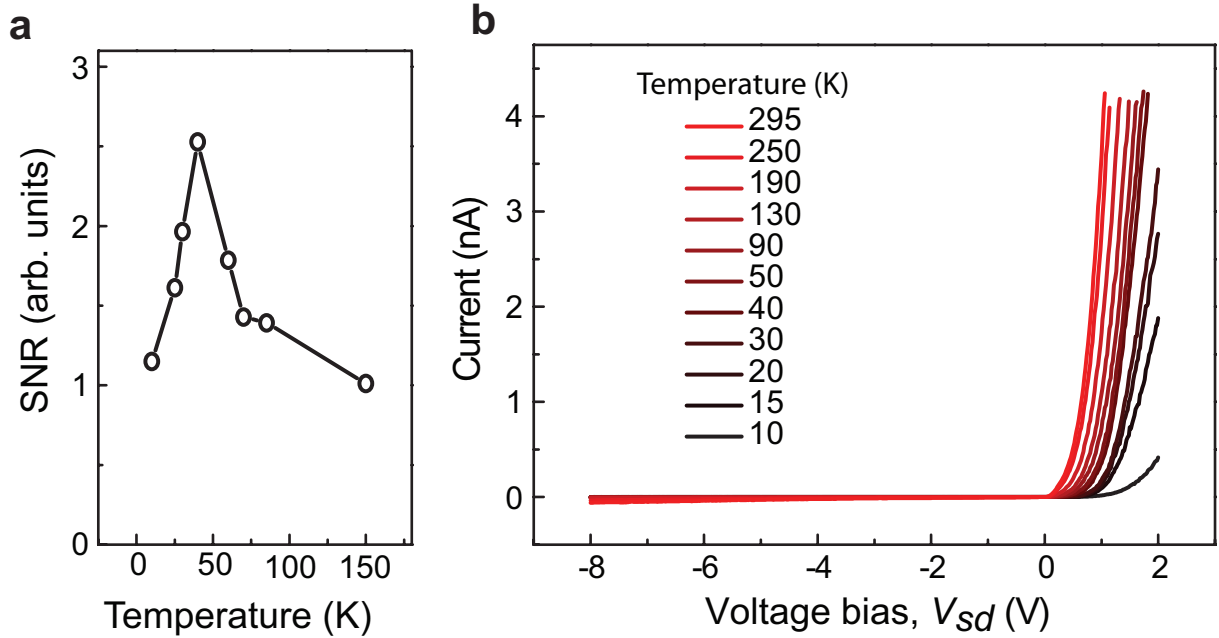}
\caption{\textbf{a}, Signal-to-Noise Ratio (SNR) of the nanowire photodiode as a function of temperature measured for an optical power of 5.6 fW incident on the sample. SNR exhibits a maximum of 2.5 at 40 K, this temperature has been used for sensitivity measurements. \textbf{b}, Electrical characteristics of the photodiode in dark as a function of the operating temperature from 295 K to 10 K. Cooling down the photodiode below 40 K produces a dramatic decrease of the forward current, which can be a result of dopant freezing in the nanowire. This effect can explain the increase noise obtained when cooling down the device below 40 K, which results in a reduction of the SNR in \textbf{a}.}

\end{figure}

\clearpage
\begin{figure}[htb]
\centering
	\includegraphics{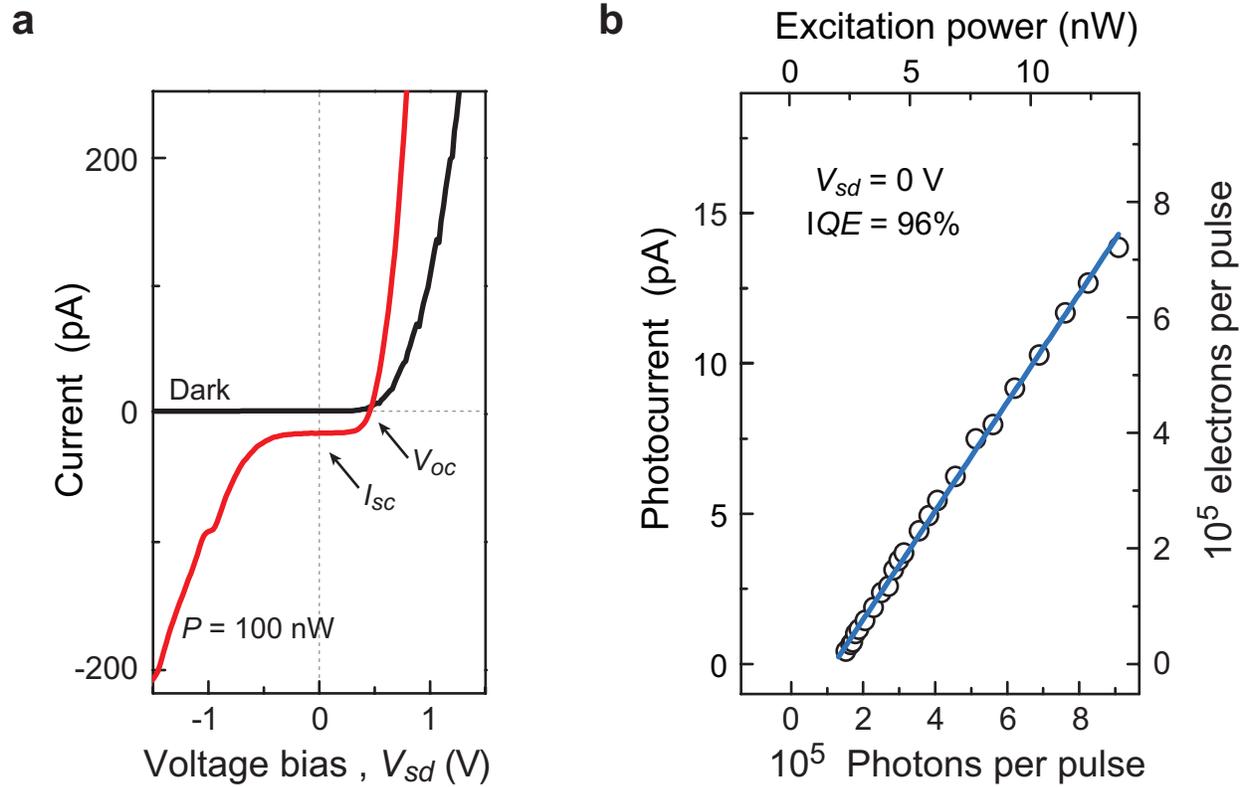}
\caption{Internal quantum efficiency of the nanowire photodiode. \textbf{a}, Current-voltage characteristic in dark (black) and under optical excitation at $\lambda\,=\,$532\,nm. We observe an open-circuit voltage $V_{oc}$\,=\,0.5\,V and photocurrent at $V_{sd}$\,=\,0\,V, which is only determined by the p-n junction built-in field without multiplication gain. \textbf{b}, From the linear fit (blue line) to experimental data (circles), we determine an efficiency of 96\,$\%$ in the photon to charge conversion and subsequent collection at the contacts.}

\end{figure}
\clearpage
\begin{figure}[htb]
\centering
	\includegraphics{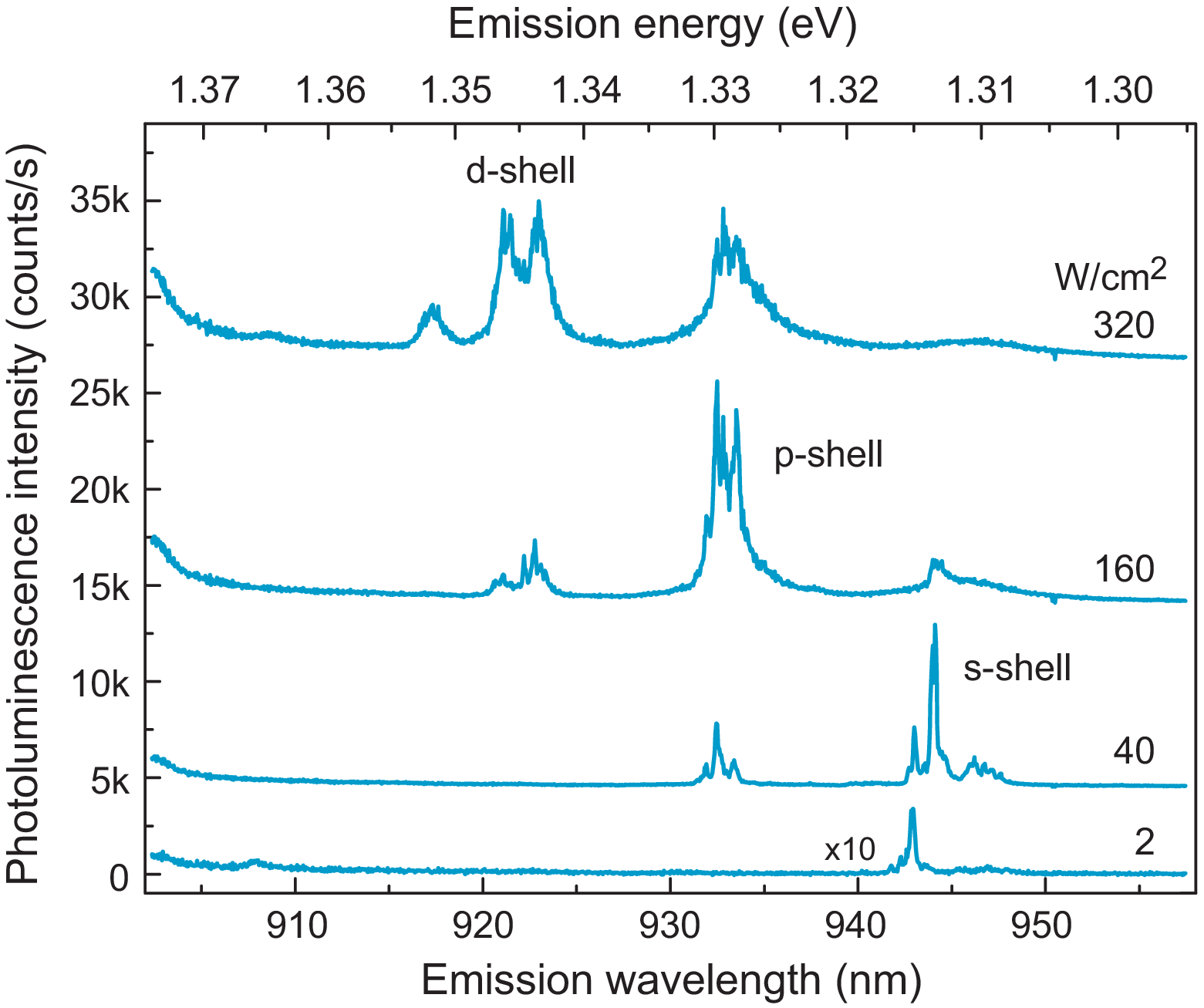}
\caption{Photoluminescence spectroscopy of a single InAsP quantum dot in an InP nanowire. Typical photoluminescence spectra as a function of laser excitation power exhibit the sequential filling of s-, p-, and d-shell as the laser excitation power is increased. A constant energy splitting is observed between the quantum dot shells.}

\end{figure}


\clearpage

\begin{thebibliography}{30}

\bibitem{Zrenner2002}
A.~Zrenner, et al.
\newblock Coherent properties of a two-level system based on a quantum-dot
  photodiode.
\newblock {\em Nature}, \textbf{418}, 612--614, (2002).

\bibitem{Koppens2006}
F.~H.~L.~Koppens et al.
\newblock Driven coherent oscillations of a single electron spin in a quantum
	dot.
\newblock {\em Nature}, \textbf{442}, 766--771, (2006).

\bibitem{Press2008}
D.~Press, T.~D. Ladd, B.~Zhang, and Y.~Yamamoto.
\newblock Complete quantum control of a single quantum dot spin using ultrafast
  optical pulses.
\newblock {\em Nature}, \textbf{456}, 218--221, (2008).

\bibitem{Brunner2009}
D.~Brunner, et al.
\newblock A coherent single-hole spin in a semiconductor.
\newblock {\em Science}, \textbf{325}, 70--72, (2009).

\bibitem{ZrennerNatPhot}
S.~Michaelis~de Vasconcellos, S.~Gordon, M.~Bichler, T.~Meier, and A.~Zrenner.
\newblock Coherent control of a single exciton qubit by optoelectronic
  manipulation.
\newblock {\em Nature Photon.}, \textbf{4}, 545--548, (2010).

\bibitem{Nadj2010}
S.~Nadj-Perge, S.~M. Frolov, E.~P. A.~M. Bakkers, and L.~P. Kouwenhoven.
\newblock Spin-orbit qubit in a semiconductor nanowire.
\newblock {\em Nature}, \textbf{468}, 1084--1087, (2010).

\bibitem{Benny2011}
Y.~Benny, et al.
\newblock Coherent optical writing and reading of the exciton spin state in
  single quantum dots.
\newblock {\em Phys. Rev. Lett.}, \textbf{106}, 040504, (2011).

\bibitem{Loss}
D.~Loss and D.~P.~DiVincenzo.
\newblock Quantum computation with quantum dots.
\newblock {\em Phys. Rev. A}, \textbf{57}, 120--126, (1998).

\bibitem{Poem2010}
E.~Poem, et al.
\newblock Accessing the dark exciton with light.
\newblock {\em Nature Phys.}, \textbf{6}, 993--997, (2010).

\bibitem{Vrijen2001}
R.~Vrijen and E.~Yablonovitch.
\newblock A spin-coherent semiconductor photo-detector for quantum
  communication.
\newblock {\em Physica E}, \textbf{10}, 569--575, (2001).

\bibitem{Kosaka2009}
H.~Kosaka, et al.
\newblock Spin state tomography of optically injected electrons in a
  semiconductor.
\newblock {\em Nature}, \textbf{457}, 702--705, (2009).

\bibitem{LieberNL}
C.~Yang, C.~J.~Barrelet, F.~Capasso, and C.~M.~Lieber.
\newblock Single p-type/intrinsic/n-type silicon nanowires as nanoscale
  avalanche photodetectors.
\newblock {\em Nano Lett.}, \textbf{6}, 2929--2934, (2006).

\bibitem{Hayden2006}
O.~Hayden, R.~Agarwal, and C.~M.~Lieber.
\newblock Nanoscale avalanche photodiodes for highly sensitive and spatially
  resolved photon detection.
\newblock {\em Nature Mater.}, \textbf{5}, 352--356, (2006).

\bibitem{Michael}
M.~E.~Reimer, et al.
\newblock Single photon emission and detection at the nanoscale utilizing
  semiconductor nanowires.
\newblock {\em J. Nanophoton.}, \textbf{5}, 053502, (2011).

\bibitem{Capasso1987}
F.~Capasso.
\newblock Band-gap engineering: From physics and materials to new semiconductor
  devices.
\newblock {\em Science}, \textbf{235}, 172--176, (1987).

\bibitem{Thomas2010}
O.~Thomas, Z.~L. Yuan, J.~F. Dynes, A.~W. Sharpe, and A.~J. Shields.
\newblock Efficient photon number detection with silicon avalanche photodiodes.
\newblock {\em Appl. Phys. Lett.}, \textbf{97}, 031102, (2010).

\bibitem{Gabor2009}
N.~M.~ Gabor, Z.~Zhong, K.~Bosnick, J.~Park, and P.~L.~McEuen.
\newblock Extremely efficient multiple electron-hole pair generation in carbon
  nanotube photodiodes.
\newblock {\em Science}, \textbf{325}, 1367--1371, (2009).

\bibitem{Tomioka2010}
K.~Tomioka, J.~Motohisa, S.~Hara, K.~Hiruma, and T.~Fukui.
\newblock GaAs/AlGaAs core multishell nanowire-based light-emitting diodes on
  Si.
\newblock {\em Nano Lett.}, \textbf{10}, 1639--1644, (2010).

\bibitem{Magnus2011}
M.~Heurlin, et al.
\newblock Axial InP nanowire tandem junction grown on a silicon substrate.
\newblock {\em Nano Lett.}, \textbf{11}, 2028--2031, (2011).

\bibitem{Messing2011}
M.~E.~Messing, et al.
\newblock Growth of straight InAs-on-GaAs nanowire heterostructures.
\newblock {\em Nano Lett.}, \textbf{11}, 3899--3905, (2011).

\bibitem{MinotNL}
E.~D.~Minot, et al.
\newblock Single quantum dot nanowire LEDs.
\newblock {\em Nano Lett.}, \textbf{7}, 367--371, (2007).

\bibitem{Pearsall1979}
T.~P. Pearsall.
\newblock Threshold energies for impact ionization by electrons and holes in
  InP.
\newblock {\em Appl. Phys. Lett.}, \textbf{35}, 168--170, (1979).

\bibitem{Algra2008}
R.~E.~ Algra, et al.
\newblock Twinning superlattices in indium phosphide nanowires.
\newblock {\em Nature}, \textbf{456}, 369--372, (2008).

\bibitem{Reimer2011}
M.~E. Reimer et al.
\newblock Bright single photon sources in bottom-up tailored nanowires.
\newblock {\em Submitted to Nat. Commun.}.

\bibitem{Atwater2010}
M.~D.~ Kelzenberg, et al.
\newblock Enhanced absorption and carrier collection in Si wire arrays for
  photovoltaic applications.
\newblock {\em Nature Mater.}, \textbf{9}, 239--244, (2010).

\bibitem{Wang2001}
J.~ Wang, M.~S.~ Gudiksen, X.~ Duan, Y.~Cui, and C.~M.~ Lieber.
\newblock Highly polarized photoluminescence and photodetection from single
  indium phosphide nanowires.
\newblock {\em Science}, \textbf{293}, 1455--1457, (2001).

\bibitem{Kouwen2010}
M.~P. van Kouwen, et al.
\newblock Single quantum dot nanowire photodetectors.
\newblock {\em Appl. Phys. Lett.}, \textbf{97}, 113108, (2010).

\bibitem{Curto2010}
A.~G.~ Curto, et al.
\newblock Unidirectional emission of a quantum dot coupled to a nanoantenna.
\newblock {\em Science}, \textbf{329}, 930--933, (2010).

\bibitem{Lee2011}
G.~K. Lee, et al.
\newblock A planar dielectric antenna for directional single-photon emission
  and near-unity collection efficiency.
\newblock {\em Nature Photon.}, \textbf{5}, 166--169, (2011).

\bibitem{AspnesPRL}
D.~E. Aspnes and A.~A. Studna.
\newblock Dielectric functions and optical parameters of Si, Ge, GaP, GaAs,
  GaSb, InP, InAs, and InSb from 1.5 to 6.0 eV.
\newblock {\em Phys. Rev. B}, \textbf{27}, 985, (1983).

\end{thebibliography}

\begin{thebibliography}{1}

\bibitem{Siegel}
W.~Siegel, G.~K\"{u}hnel, H.~Koi, W.~Geelach.
\newblock Electrical properties of n-type and p-type InP grown by the synthesis, solute diffusion technique.
\newblock {\em Phys. Stat. Sol. (a)}, \textbf{95}, 309, (1986).

\end{thebibliography}
\end{document}